# INFRARED COUNTERPARTS OF X-RAY GALAXIES

A.V. Tugay[1], S.Yu.Shevchenko[2]
[1]Taras Shevchenko National University of Kyiv, Kyiv, Ukraine,
tugay.anatoliy@gmail.com
[2]Schmalhausen Institute of Zoology, NASc of Ukraine, Kyiv, Ukraine,
astromott@gmail.com

Population studies of the extragalactic objects are a major part of the universe large-scale structure study. Apart from radio, infrared, and visible wavelength bands, observations and further identification of extragalactic objects such as galaxies, quasars, blazers, liners, and active star burst regions are also conducted in the X-ray and gamma bands. In this paper we make identification and cross-correlate of the infrared and X-ray observational data, build a distribution of a selected sample sources by types and attempted to analyze types of the extragalactic objects at distances up to $z = 0.1$ using observational data of relevant space observatories.

Data from a leading X-ray space observatory XMM-Newton were used to compile the largest catalog of X-ray sources. Current version of XMM SSC (Serendipitous Source Catalog) contains more than half a million sources. In our previous works we selected and analyzed a sample of 5021 X-ray galaxies observed by XMM-Newton. Identification and classification of these sources is essential next step of the study. In this study we used infrared apparent magnitudes from WISE catalog of AGN candidates. In 2010 space telescope WISE performed full sky survey in four infrared bands and detected 747 million sources. WISE catalog of AGN candidates amounts 4 million of possible extragalactic sources. We built infrared color-color diagram for our sample of X-ray galaxies and assessed their types using WISE telescope data. In this study we also analyzed large scale structure of the universe (distances up to $z=0.1$). This analysis revealed Coma galaxy cluster and SDSS Sloan Great Wall. In the further studies we are planning to investigate the distribution of different types of X-ray galaxies within the large-scale structures of the Universe.

Інфрачервоні відповідники рентгенівських галактик

Тугай А.В., Шевченко С.Ю.

Дослідження складу і природи позагалактичних об'єктів є важливою частиною досліджень великомасштабної структури всесвіту. Окрім радіо, інфрачервоного діапазону та видимого випромінення, спостереження та подальше ототожнення позагалактичних об'єктів, таких як скупчення галактик, квазари, блазари, лайнери та області з активним зореутворенням ведуться в рентгенівському та гама діапазонах. В цій роботі ми провели ототожнення та перехресну кореляцію спостережних даних інфрачервоного та рентгенівських діапазонів та побудували розподіл в вибірці цих джерел за типом. Проаналізовано вміст позагалактичних об'єктів різних типів на відстанях до z=0.1 використовуючи дані відповідних космічних обсерваторій.

Для складання найбільших каталогів рентгенівських джерел використовуються дані провідної рентгенівської космічної обсерваторії XMM-Newton. Поточна версія каталогу XMM SSC (Serendipitous Source Catalog) містить понад півмільйона джерел. У наших попередніх роботах ми відібрали та проаналізували дані 5021 галактики в рентгенівському діапазоні, які спостерігались XMM-Newton. Ідентифікація та класифікація цих джерел є важливим наступним кроком дослідження. У цьому дослідженні ми використали інфрачервоні зображення інфрачервоної космічної обсерваторії WISE з каталогу кандидатів у активні ядра галактик. У 2010 році космічний телескоп WISE здійснив огляд всього повного неба у чотирьох інфрачервоних діапазонах та виявив 747 мільйонів джерел. Каталог WISE кандидатів у активні ядра галактик становить 4 мільйони можливих позагалактичних джерел. Ми побудували інфрачервону кольорову діаграму для нашої вибірки рентгенівських галактик та оцінили їх типи, використовуючи дані телескопа WISE. В цій роботі ми також дослідили великомасштабну структуру Всесвіту (на відстанях до z=0.1). Цей аналіз виявив скупчення галактик кластер Coma та Велику стіну Слоана (Sloan Great Wall). У подальших дослідженнях ми плануємо дослідити розподіл різних типів рентгенівських галактик у великомасштабних структурах Всесвіту.

Ключові слова: Активні ядра галактик, XMM-Newton, WISE, рентгенівське випромінювання, інфрачервоні огляди.

1. Introduction

Multiwavelength observations have a great importance for extragalactic astronomy. Many manifestations of the galactic activity could be detected in X-ray and infrared bands. In this respect the most vast database of X-ray sources is contained in XMM-Newton Serendipitous Source Catalogue (XMM-SSC). Current version of XMM-SSC, 3XMM-DR8 comprises more than 500 thousand sources (Rosen et al., 2016). We compiled a sample of extragalactic sources by previous version of XMM-SSC - 2XMM (Watson et al., 2009). Our sample, herein Xgal, contains 5021 entries (Tugay, 2014). HyperLeda database (http://leda.univ-lyon1.fr, Makarov et al., 2014) was used for identification extragalactic X-ray sources and further Xgal compilation. Classification of Xgal sources is the next objective of our study. Most of X-ray emitting galaxies are more likely to have AGNs (Tugay & Vasylenko, 2011), but this statement should be verified.

For the purpose of this study we used data from the sky survey performed by Wide-field Infrared Survey Explorer (WISE). This space observatory registered more than 747 million infrared emission sources. The WISE telescope has a 40-centimeter-diameter aperture. Its' four working wavelength bands are 3.4, 4.6, 12 and 22 microns. The main astronomical sources irradiating in such bands are the following:

W1 band (3.4 micrometers) – stars and galaxies;
W2 band (4.6 micrometers) – thermal radiation from the internal heat sources of sub-stellar objects like brown dwarfs;
W3 band (12 micrometers) – thermal radiation of asteroids;
W4 band (22 micrometers) – dust in star-forming regions.

The first three band are often used to build color diagrams for classification of WISE sources. Distribution of the different cosmic objects by types and their IR colors is presented on Fig. 1
 (http://wise2.ipac.caltech.edu/docs/release/allsky/expsup/index.html, Cutri et al., 2013).
The purpose of the current work was to identify Xgal sources in WISE catalog and classify them using established infrared colors.

2. Sample Studies

One of the study objectives was to perform cross identification between WISE all sky observations (Cutri et al., 2013; 747.634.026 sources) and X-ray sources observation using already compiled catalogues: XMM-Newton Serendipitous Source Catalogue 2XMMi-DR3 (262.902 sources); HyperLeda (2.777.804 galaxies) and Xgal (Tugay, 2012; 5.021 sources). Some previously obtained multiwavelength cross-correlation results based on data from WISE, 3XMM, and FIRST/NVSS for 2529 sources are given in paper of Mingo et al. (2016). In our study we selected the most bright X-ray galaxies with a flux $F_X>10^{-13}$ mW/m$^2$. Classification of these objects was performed using SIMBAD Astronomical Database.

Initially we assumed that Xgal objects should be in WISE AGN catalog (Assef et al., 2018), since as it was mentioned above, most of X-ray galaxies are AGNs. Our results are given below on Fig. 2. We identified 267 sources with W1-W2>0.5 that corresponds to QSO region on WISE color diagram (Fig. 1). WISE AGN catalog was compiled using mentioned selection criterion and contains approximately 4 million infrared AGN candidates.

Then we performed cross correlation of Xgal with a whole AllWISE catalog (Cutri et al., 2013) and found 4 thousands matching objects that roughly corresponds to the number of X-ray WISE sources in (Mingo et al., 2016). We have split SIMBAD types into three groups: ellipticals, spirals and AGNs. Distribution of bright Xgal sources of these three types was plotted on Fig. 3.

On this chart it is possible to distinguish quasars (color index W1-W2>0.5) and galaxies (color index W1-W2<0.5) which could be divided into elliptical and spiral types. Elliptical galaxies emit X-rays from cluster or group halo. Spiral galaxies may have X-ray emission from star formation regions (Tugay & Vasylenko, 2011). Such galaxies are not numerous in our sample. Moreover, the following two comments could be made here. First, in MIXR sample (Mingo et al., 2016) there is a division for spiral and star-forming galaxies. We can not make such a difference in our sample. Second, a subset of the galaxies that we mark as 'spirals' has an incorrect position on a color diagram. The explanation to this could be that some of the galaxies classified in SIMBAD as spirals are in fact Seyfert galaxies and they occupy the upper part of the plot on Fig. 3. Lower part of the 'spirals' chart indicates that there should be an X-ray halos envelope around these galaxies.

We have also analysed radial distribution of the bright X-ray galaxies in Sloan Digital Sky Survey region. In case of uniform distribution, the number of galaxies should increase as a cubic function of the distance. And vice versa, radial velocity of a galaxy $V_{3K}$ is expected to be increased as cubic root function of galaxy order number in the list sorted by velocity value. Any major deviations from uniform spatial distribution of galaxies should lead to divergence of $V_{3K}(N)$ relation from cubic root. Such deviations becomes the most visible in our sample after replacing $V_{3K}(N)$ by $V_{3K}/1.007^N(N)$ (Fig. 4; power-law base factor equal to 1.007 was selected empirically). The main trend of such a function for the main part of the sample is a horizontal line approximation. Additional increase of this function corresponds to larger distances between galaxies, e.g. voids. The declination on the graph corresponds to dense large-scale structures (LSS) in galaxy distribution. Coma galaxy cluster and Sloan Great Wall were revealed in this distribution. Behind Sloan Great Wall the number of X-ray galaxies becomes insignificant that makes LSS detection impossible and causes rapid growth of plot on Fig. 4 for largest N.

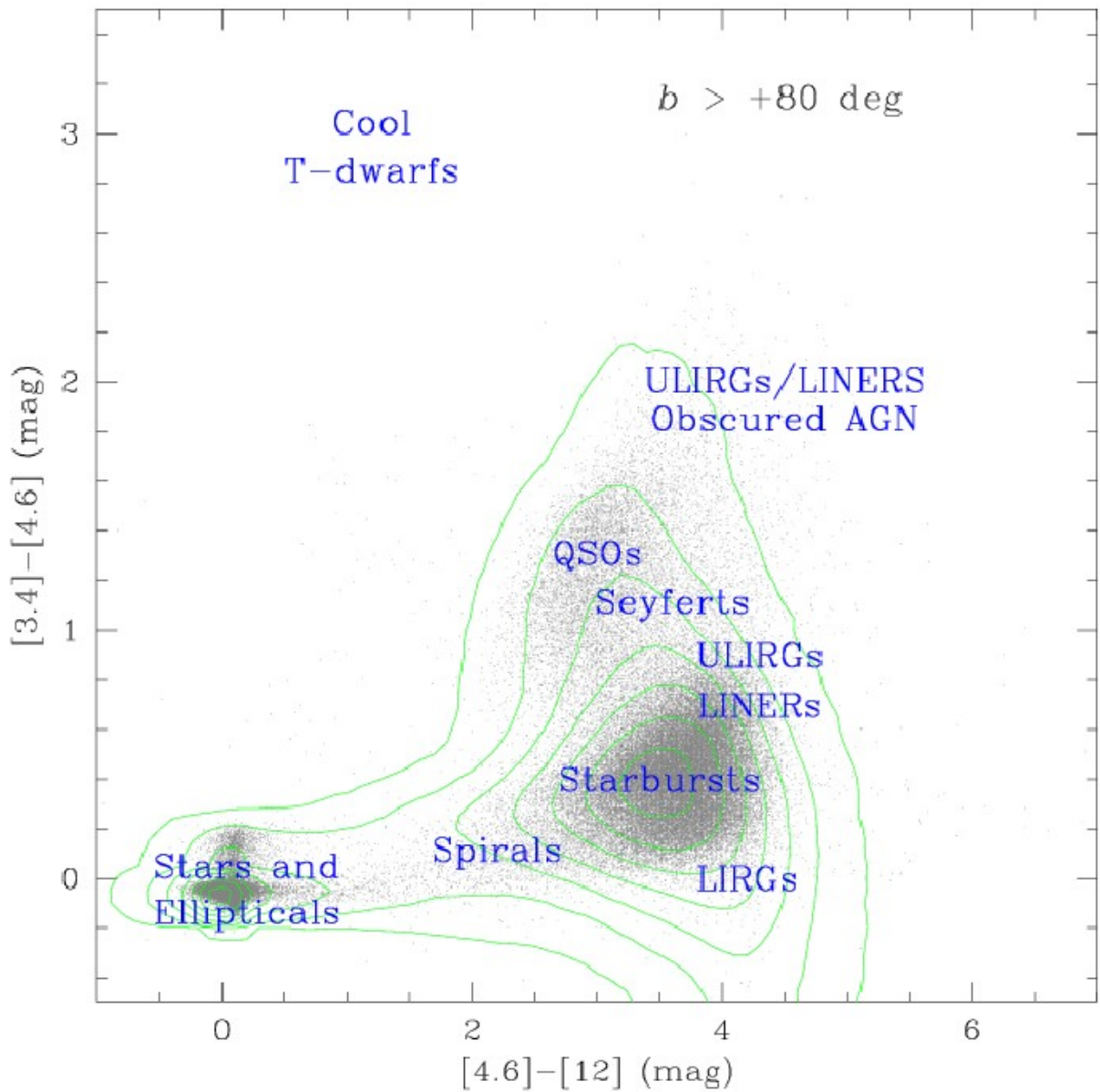

Figure 1. IR wave bands and objects types observed by Wide-field Infrared Survey Explorer (Cutri et al., 2013)

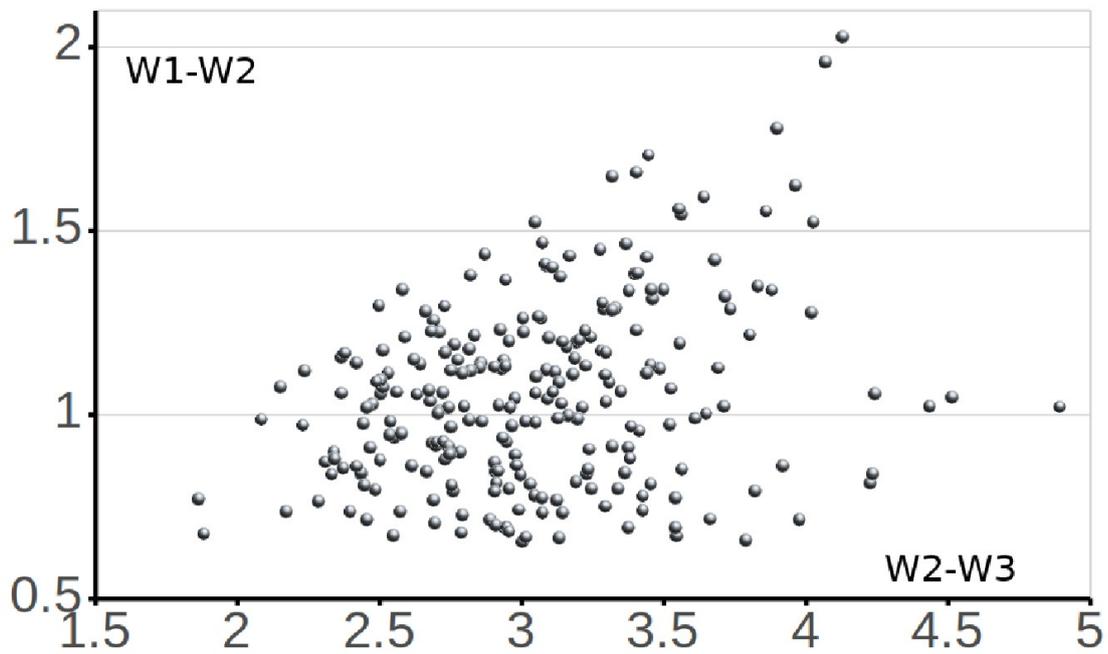

Figure 2. WISE color diagram for AGN candidates from (Assef et al., 2018) found in Xgal sample.

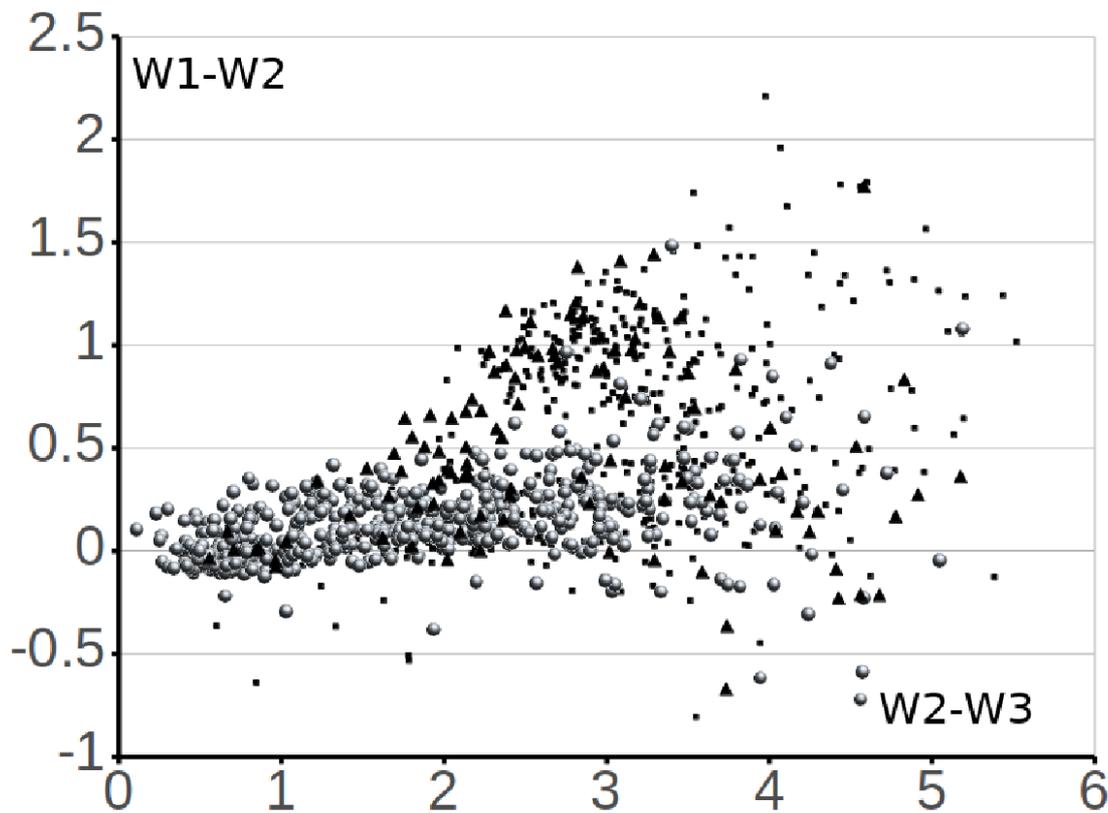

Figure 3. The MIXR sample: AGN activity versus star formation. IR color diagram for 573 X-ray galaxies with $F_X>10^{-13}$ mW/s/m$^2$. Balls — elliptical galaxies, galaxies in groups and clusters. Triangles — spiral and star-forming galaxies. Dots — AGN's : QSO's, Seyferts, LINERS

3. Conclusion

Elliptical and clustered X-ray galaxies has a clear stand out position on the infrared color-color diagram. Main source of their X-ray emission should be explained as interstellar and intergalactic gas. Quasars also could be distinguished with WISE magnitudes. Main source of X-ray emission AGNs is accretion disc around central black hole. Star-forming galaxies has neither clear region at color diagram nor significant representation among X-bright galaxies. Some LSS elements could be resolved in radial distribution of X-ray galaxies. Such studies could be developed and combined with other methods of LSS analysis (Tugay, 2014; Tugay et al., 2016).

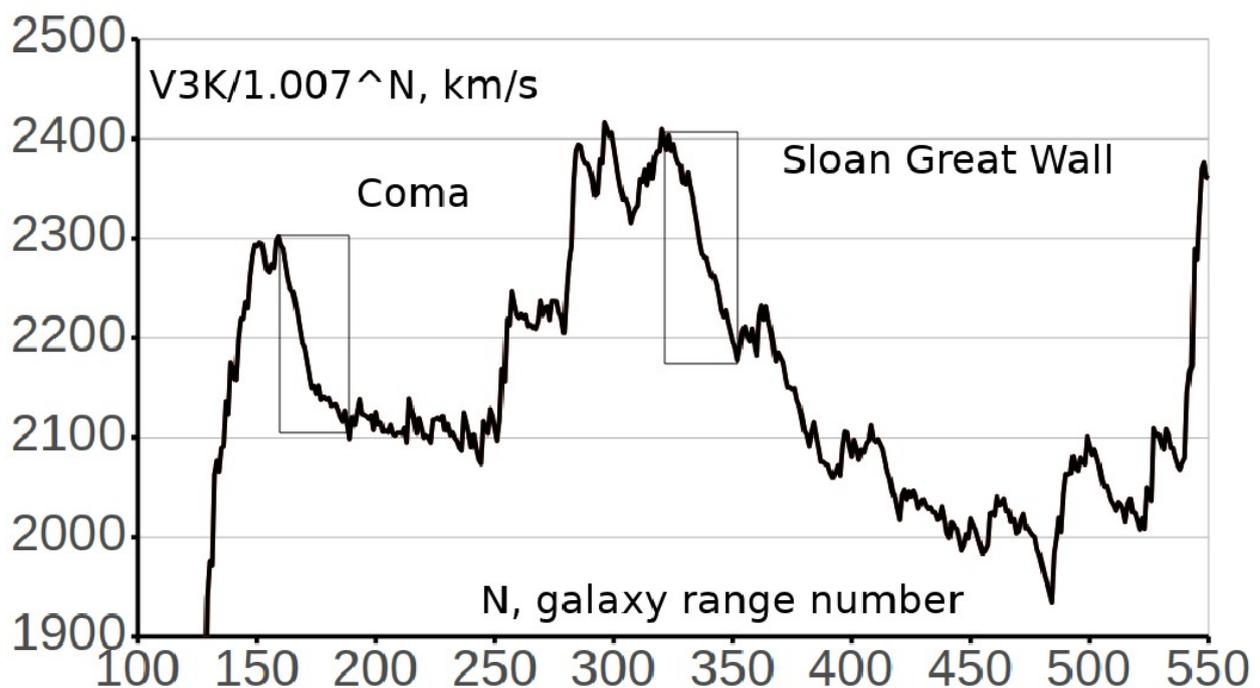

Figure 4. Large-scale structures in radial distribution of X-ray galaxies

Acknowledgements.

This study was conducted using the data from the 3XMM XMM-Newton serendipitous source catalogue compiled by the 10 institutes of the XMM-Newton Survey Science Centre selected by ESA. The authors are thankful to Dr. L.Stawarz for scientific objective targeting and consultations.
We acknowledge the usage of the HyperLeda database (http://leda.univ-lyon1.fr).; and SIMBAD Astronomical Database - CDS (Strasbourg)
 http://simbad.u-strasbg.fr/simbad/